\documentclass[aps,twocolumn,10pt]{revtex4}
\usepackage{amsfonts}
\usepackage{amsmath}
\usepackage{amssymb}
\usepackage{graphicx}
\usepackage{epsfig}

\begin{document}

\title{Imaging of critical correlations in optical lattices and atomic traps}
\author{Qian Niu$^{a)}$, Iacopo Carusotto$^{b)}$, and A. B. Kuklov$^{c)}$ }
\affiliation{$^{a)}$Department of Physics, University of Texas, Austin, TX 78712\\
$^{b)}$BEC-CNR-INFM and Dipartimento di Fisica, Universita di Trento, I-38050 Povo, Italy\\
$^{c)}$Department of Physics,CSI, CUNY - Staten Island, New York, NY 10314}

\begin{abstract}
We consider critical space-time correlations in trapped utracold atoms and propose a method
for their detection by the interference with a reference Bose-Einstein condensate. An important
information about universal properties can be obtained despite non-uniformity and small
sizes of atomic clouds.  
The advantage of this method is that
the correlation properties of the interference pattern are robust with respect
to expansion and are not, practically, affected by microscopic structure of underlying lattice including
structural disorder.
The non-destructive scheme allows measuring the space-time correlators of arbitrary order.  
\end{abstract}

\maketitle

\section{Introduction}
Experimental realization of the Mott insulator (MI) superfluid (SF)
transition in optical lattices \cite{MI-SF} opened new
directions in studying strongly correlated states of matter.
Mott insulator of single-component bosons represents a simple and
yet very rich strongly interacting system.
Several other strongly correlated phases have also been predicted recently
\cite{Zhou,SCF,We,Demler}. The impact of disorder on the transition is another area of great interest
\cite{Boseglass}, which can also be addressed in optical lattices (OL) \cite{glass_exp}. Despite these quite exciting
developments, the fundamental aspect of the many-body physics --- critical behavior upon approaching a continuous transition point --- evades experimental determination in optical lattices.

The critical behavior is characterized by the divergence of the correlation length $\xi$ as the system approaches the critical point from either side of the phase transition.  On the MI side, $\xi$ can also be viewed as the coherence length controlling exponential decay of the phase correlation.  On the SF side, the system acquires a global average phase so that the coherence length is determined by the system size $L$. In this case, the correlation length $\xi$ gives the length scale of fluctuations on top of the established off-diagonal long range order.  Thus, $\xi$ is finite and  coincides with the healing length (the length scale on which the order parameter responds to a local perturbation).  This divergence of $\xi$ is the key feature of the hyperscaling paradigm of the continuous phase transitions. The exponent $\nu$ characterizes the singular growth of $\xi$ as the system is tuned into the critical point and uniquely represents the universality class of the transition.       

Coherence properties of Bose-Einstein condensate (BEC) have been demonstrated in the seminal experiments \cite{Ketterle,note,Bloch1}. In contrast, no  systematic studies of the critical properties were reported so far.
Relatively small and non-uniform samples as well as the specificity of the imaging methods do not allow direct measurement of divergent $\xi$.  
As discussed in ref.\cite{Damle}, a presence of the parabolic trapping potential requires fulfilling
special conditions in order to observe critical behavior. For some quantities even critical
exponents can be modified because of the trapping potential. In recent Monte Carlo studies \cite{troy2,troy},
it has been shown that practical realization of the critical behavior requires suppression of the parabolic
potential and using, instead, quartic traps so that the central region of highest density can well
be approximated by uniform box which is the standard assumption of the finite size scaling approach (FSS) (see in Ref.\cite{FSS}) developed for macroscopic and uniform (apart from sharp boundaries) samples traditionally studied so far. Thus, consistent application of the FSS paradigm in the cases of trapping potential requires separate study. 
In the present paper, we will focus our attention on the problem of
  measuring the correlators of the atomic field in OL. Clearly, a solution of this problem is a necessary
prerequisite for a consistent study of the criticality in OL and traps.
 For sake of concreteness, we shall ignore the issue
of the shape of the trapping potential \cite{Damle,troy2,troy} and follow the standard FSS approach \cite{FSS}.  

The absorptive imaging techniques used in, e.g., Ref.\cite{imaging} in their present form
do not leave much room for improvement. Indeed, the far zone imaging of the cloud released from an optical lattice for the observation of MI-SF transition relies on analyzing the Bragg interference pattern. The information on the superfluid off-diagonal correlation length is encoded in the width of the Bragg peaks \cite{Borya_Kolya}. This information can easily be distorted due to many external causes. Furthermore, this imaging method becomes virtually useless for the study of superfluidity in the presence of purposely imposed disorder \cite{glass_exp}, because rather the lattice disorder than the subtle many-body correlations will immediately broaden the Bragg peaks. Thus, it is very desirable to have a method which will allow a real space imaging of the long-range off-diagonal correlations regardless of the underlying structure of the optical lattice.

We propose such a method of {\it reference interference imaging } in which 
the tested cloud is allowed to interfere with some reference BEC. Here we 
consider 
the simplest case of one-component bosons undergoing MI-SF transition.
The question to what extent the proposed method can be applied to more
complex situations of several species is left out for further study.
Our idea stems from the observation that quasi-coherence in the MI phase exists on sizes $R$ such that $a_l \ll R \leq \xi$, where $a_l$ stands for the lattice constant. This view is closely related to the idea of quasi-condensate introduced, first, by Popov \cite{Popov} and further developed in refs.\cite{Kagan,Borya} with respect to weakly interacting Bose gas undergoing
a transition from a normal state to the BEC.
These two seemingly very different systems can be mapped on to each other
in the critical region with the help of the quantum-to-classical mapping \cite{Sachdev} so that the critical behavior of the quantum MI-SF transition (at temperature $T=0$) in d-spatial dimensions is equivalent to that of the Bose-Einstein condensation in D=d+1 dimensions (at finite $T$).  In other words, both systems belong to the same universality class $U(1)$ characterized by identical critical exponents.
Thus, studying the critical behavior can equally be done either in optical lattices at $T=0$ in $d=2$ geometry or in the $d=3$ traps close to the BEC-formation temperature \cite{note1}. Finding these exponents is a very challenging experimental \cite{alpha_exp} and computational task \cite{Burovski}.

It is also worth noting that both transitions are characterized by a large critical region, which makes it quite promising to study the criticality in the trapped atomic systems.  In some systems such as traditional superconductors, mean field behavior is predominant everywhere except for a very small region (as small as $10^{-15}$ of the transition temperature) about the critical point where true critical behavior prevails.   For a weakly interacting boson gas in free space, the relative width of the critical temperature interval around
the condensation temperature $T_c$ is given by the mean density $n$ and the scattering length $a$ as $\approx 10^2 n^{1/3}a$ \cite{Borya}. Thus, even for the gas parameter as small as $na^3=10^{-6}$, the critical region is comparable to $T_c$.  Similarly, in the case of the zero temperature MI-SF transition in d=2 dimensional optical lattices, the critical region width of the quantum transition is given by $\delta u/u_c\approx 1/\tilde{n}$, where $\tilde n$ is the mean occupation number per lattice site, and $u=U/J$ with $U$ and $J$ standing for the onsite interaction and the tunneling amplitude, respectively \cite{Altman}.  For $\tilde{n}=1$, the relative size of the critical region is unity.  We therefore expect that the critical relation $\xi \approx a_l/|\delta u/u_c|^\nu$ may be used to estimate the coherent length reliably down to very small length scales.  For example, in order to observe $\xi \approx 10a_l$, one needs $\delta u \approx 0.03u_c=0.7$ in d=2 square lattice.

When the correlation length becomes mesoscopic, a particular region of linear dimension $\sim \xi$ can be viewed as a domain of coherent SF. Such a finite size SF is characterized by uniformity of the classical phase. If the optical lattice is removed and the cloud is allowed to overlap with some reference BEC cloud moving coherently at momentum $k_r \gg 1/\xi$, an interference pattern will be formed. This pattern will consist of domains of typical sizes $\sim \xi$, with each domain featuring a typical interference pattern of parallel lines separated by the distance $2\pi / k_r$ \cite{Ketterle}.  Pronounced phase shift (or distortion) in the parallel lines will occur over the length scale of $\xi$.

This proposal has a very close analogy in quantum
optics ~\cite{WallsMilburn}:
Measurements of the phase of an optical beam are
generally performed by sending the beam onto one input arm of a
beam-splitter, and the (strong) reference beam onto the other input arm.
The intensity and noise in the two output arms carry information on a
given quadrature of the optical beam, and on its variance. In
particular, squeezing experiments consist in observing a reduced noise
on a quadrature. In the matter waves field the interference between two condensates has been employed
for identifying vortices \cite{vortex}.

It is important to note that the relative phase between the
  optical lattice cloud and the reference BEC varies from shot to shot
  in an uncorrelated way, so that no interference pattern will
  appear when averaged over many realizations.
However, as we will show below, the combination of the interference
with the measurements of the noise correlator of the
 Hanbury Brown \& Twiss (HBT) type \cite{HBT,Bloch_noise,Demler2,Aspect},  which gives the density-density correlator at different spatial locations, can provide the information on diverging correlations. Related methods have been recently proposed to determine
long-range order in fermionic systems~\cite{IC_YC}. Very recently, the noise analysis of the interference
between two low-dimensional condensates has been proposed for detecting the specific fluctuating behaviors \cite{Polkovnikov}.
We will also discuss that the reference interference imaging can be adopted for deriving a full space-time correlator, which carries the most complete information about a system, including excitation spectrum and the dynamical scaling.
Furthermore, the non-destructive variation of the method employing outcoupling atomic beams
allows, in principle, measuring higher order correlators -- n-body Green's functions.

\section{Correlation properties of the interference pattern}

In the following subsections, we, first, discuss the Bragg
interference of the lattice atoms alone in order to emphasize the limitations of the Bragg
method. Second, we will consider the interference
between the lattice atoms with a reference BEC so that the advantage of using the reference interference imaging
will become evident. The analysis of a simplified 
scheme, in which the reference cloud is taken as uniform,
will be followed by a discussion of a more practical setup later in Sec.\ref{prac}

\subsection{Density imaging of expanding cloud released from a lattice}

Earlier experiments focused on expanding density profile of
atoms released from a trap or a lattice, and here we give a brief discussion how
this may or may not reflect the superfluid correlations
in the lattice. In general, density profile $n({\bf x},t)$ of a cloud undergoing free expansion 
is determined by distribution of momenta carried by particles.
In other words, $n({\bf x},t)$ represents spatial Fourier transform of the
one-particle density matrix with momentum ${\bf q} = {\bf x}/t$. (Here and
below, unless stated otherwise, we employ atomic units in which
atomic mass $m$ and the Planck constant $\hbar$ are unity).

In the absence of the lattice potential, when $\xi$ is
the shortest length scale which defines the largest momentum $1/\xi$,
the expanding cloud size $R$ at long times would be uniquely
determined by $\xi$ as $R=t/\xi$.
In the presence of the lattice, higher spatial harmonics impose a
background which has nothing to do with the critical behavior and which
complicates the analysis.

Onsite packets of Wannier functions $W_t({\bf x} - {\bf x}_i)$, where ${\bf x}_i$ is the position of site $i$, contribute the fastest components into the expansion. Their expansion speed is determined by the uncertainty principle $V_w\sim 1/a_w$, where $a_w$ is the width of the initial Wannier function. 
Thus, a typical size of the packet becomes $R_w\approx t/a_w$ after a brief initial transient 
on the order of $t_w \approx a^2_w$. Expansion of each Wannier packet introduces an additional phase
factor $\exp(i{\cal A})$ given by the classical action of a free
particle ${\cal A}= ({\bf x} - {\bf x}')^2/(2t)$, where ${\bf x}'$ and ${\bf x}$ are the initial and final positions respectively. 
Therefore, the result of free expansion of a Wannier function can be represented as
\begin{equation}
W_t({\bf x} -{\bf x}_i)=\frac{1}{t^{3/2}}{\rm e}^{\frac{i({\bf x}-{\bf x}_i)^2}{2t}}
\int d{\bf x}'{\rm e}^{\frac{i({\bf x}-{\bf x}_i){\bf x}'}{t}} W_0({\bf x}').
\label{t>t_1}
\end{equation}%
Here, $W_0$ denotes the Wannier function at the initial time, and we have neglected a phase of ${\bf x}'^2/2t$ which is much smaller than unity after the transient when ${\bf x}'$ is taken within the range $a_w$ of the initial Wannier function.  The factor of $t^{-3/2}$ takes care of the decrease in the magnitude as a result of the expansion. 
 
Simplicity is achieved after a relatively long expansion time, $t/a_w>>L$, where $L$ is the linear scale of the lattice.  By now, the Wannier function has expanded much beyond the original size of the lattice.  
In this case, the phase term ${\bf x}_i {\bf x}'/t$ is  small, and one arrives at the expression traditionally used for the time of flight imaging
$W_t({\bf x} -{\bf x}_i)=t^{-3/2}\exp(i({\bf x}-{\bf
x}_i)^2/(2t)) \tilde{W}_0({\bf q})$, where $\tilde{W}_0({\bf
q})$ stands for the Fourier transform of $W_0({\bf x})$ at ${\bf
q}={\bf x}/t$. 
The expression (\ref{t>t_1}) for the evolution of
the Wannier function can be employed in the field operator for the lattice atoms
\begin{equation}
\psi_l({\bf x},t)=\sum_i W_t({\bf x}-{\bf x}_i)\,b_i,
\label{psi_l}
\end{equation}
where $b_i$ is the destruction operator for the Wannier state on the $i$th site of the lattice.
Apart
from an unimportant numerical factor, the imaged density $n({\bf
x},t)=\langle \psi_l^\dagger ({\bf x},t)\psi_l({\bf x},t)\rangle$
becomes
\begin{eqnarray}
n({\bf x},t)&=&\frac{1}{t^3}|\tilde{W}_0({\bf q})|^2S({\bf q}),
\label{n(r)} \\
S({\bf q})&=&
\sum_{ij}{\rm e}^{i{\bf q}({\bf x}_i -{\bf x}_j)}
{\rm e}^{i\frac{{\bf x}^2_i -{\bf x}^2_j}{2t}}\langle b^\dagger_j b_i\rangle.
\label{S}
\end{eqnarray}%

Further simplification can be achieved in the far zone, when $t>> L \xi$.  
The phase factor ${\rm e}^{i\frac{{\bf x}^2_i -{\bf x}^2_j}{2t}}$ can be
then ignored.  It allows to relate $n({\bf r}) $ directly to the Fourier transform of the lattice correlator $S({\bf q})=\sum_{ij}\exp(i{\bf q}({\bf r}_i - {\bf r}_j))\langle b^\dagger_ib_j\rangle$.
This expression indicates that, as long as the OL is regular enough, the Bragg structure of peaks will be seen, with their width determined by $1/\xi$ (or eventually limited by finite size effects in the critical regime where $\xi>L$).  
However, if the lattice is not regular, the correlations in $\langle b^\dagger_ib_j\rangle$ will be strongly obscured by the factor $\exp(i{\bf q}({\bf r}_i - {\bf r}_j))$.  In what follows, we will show that this limitation can be overcome by measuring the interference pattern with a reference BEC.

\subsection{Interference with a reference BEC}\label{IB}

For illustration, we imagine that immediately after the lattice
was instantaneously removed, a coherent matter plane wave of the form $\exp(i{\bf k}_r{\bf x})/\sqrt{\Omega}$, 
normalized by some large volume $\Omega$,
is sent onto the lattice and is made superimpose with the lattice atoms
(more details are given in Sec.III).
The result of a single-shot absorptive imaging is given by the state average
of the total density
$n({\bf x})=\langle S| \psi^\dagger({\bf x})\psi({\bf x}) |S\rangle$,
with the field operator represented as
\begin{equation}
\psi ({\bf x})=\sum_i W_t({\bf x} - {\bf x}_i) b_i +  \frac{{\rm e}^{i{\bf k}_r{\bf x}}}{\sqrt{\Omega}}b_{rf},
\label{total}
\end{equation}%
where $b_{rf}$ is the destruction operator in the reference BEC state. 
The total density contains an interference term proportional to 
\begin{equation}
\langle S| b^\dagger_{rf} b_i |S \rangle \approx \sqrt{N_{rf} \tilde{n}}\, {\rm e}^{i\varphi_i},
\label{phase}
\end{equation}
where $N_{rf}$ stands for the total number of atoms in the reference cloud, whereas $\tilde{n}$ and $\varphi_i$ characterize the coherent nature of the lattice atoms in the state $|S\rangle$.  

According to the theory of critical phenomena, the destruction of the SF phase occurs because of large fluctuations in the phase $\varphi_i$ while the amplitude $\tilde{n}$ remains relatively constant.  For this reason, one can talk about a quasi-condensate for the lattice atoms through the critical region of the phase transition.  One may view $\tilde n$ as the quasi-condensate density per site, and $\varphi_i$ as a classical phase with the same significance as that of a classical wave.  Indeed, for $\xi^d \tilde{n} \gg 1$, the state $|S\rangle$ can be represented in the coherent state picture specified by these parameters \cite{YC_JD}.  

Although these phases vary randomly from shot to shot, they may be regarded as fixed in a single shot interference measurement \cite{Ketterle}.  Spatial dependence of the classical phase contains crucial information about the state of the system.  Deep in the SF phase, $\varphi_i$ is uniform in space.  Upon approaching the critical point, it fluctuates but retains a uniform part, which becomes zero at the critical point.  On the MI side of the phase transition, the uniform part of the phase does not exist anymore in the sense that the spatial average of $\exp(i \varphi_i)$ vanishes in thermodynamical limit.  The length scale of phase fluctuations is the correlation length $\xi$, which diverges upon approaching the critical point from either side of the phase transition.   

How can one observe the spatial behavior of the classical phase? 
The interference fringes are given by the cross-term in the density
\begin{eqnarray}
\delta n ({\bf x})=A({\bf x})\,{\rm e}^{-i{\bf k}_r{\bf x}}   + c.c.,
\nonumber \\
A({\bf x})=\sqrt{n_{rf} \tilde{n}} \sum_j W_0({\bf x} - {\bf x}_j){\rm e}^{i\varphi_j},
\label{Amp}
\end{eqnarray}%
where $n_{rf}=N_{rf}/\Omega $ is the reference cloud density \cite{noteniu}.  
Assuming that the correlation length is much larger than the lattice
constant, we can replace $\varphi_j$ by $\varphi(\bf x)$ in the expression of the amplitude Eq(\ref{Amp}), where $\varphi(\bf x)$ is a smooth interpolation of the classical phases on the lattice sites.  The interference fringes are then described by a sinusoidal function of ${\bf k}_r{\bf x}- \varphi(\bf x)$ with a modulating amplitude given by the magnitude of $\sqrt{n_{rf} \tilde{n}} \sum_j W_0({\bf x} - {\bf x}_j)$. 
Such amplitude exhibits strong spatial oscillations on the length scale
of the lattice constant. Already at expansion times $\xi^2 \gg t \gg L
a_w$, the fast oscillations have almost completely been smoothened out
(what remains of them, can still be filtered out by means of Fourier
analysis). At the same time the phase domains of the size $\approx \xi$ are still well separated from each
other and feature the interference pattern of parallel stripes
separated by a distance $2\pi /k_r$.

In practice, the role of larger expansion of the lattice atoms after the
release of the optical lattice must be considered, because it is
difficult to carry out the measurement in the short time before
expansion.  However, the expansion does not change the correlation
length. Elimination of the OL transforms the cloud into a weakly
interacting system so that the subsequent expansion can be
considered essentially free. Indeed, the correlation properties
are determined by a typical width of the momentum distribution in
the system. Release of the OL essentially freezes in this
distribution, which implies "conservation" of $\xi$. Therefore,
the qualitative features discussed above still apply to the
interference pattern between the expanded lattice-atom cloud and
the reference BEC.  

After smoothening out the short range modulations, the interference pattern consists of parallel stripes separated by a distance determined by the wavevector ${\bf k}_r$.  Spatial variation of the classical phase gives rise to distortion of the stripe positions on the scale of the correlation length.   
The stripe pattern is visible in a single-shot measurement provided each domain contains a large number ${\tilde n}\,\xi^d\gg 1$ of atoms so as to overcome shot noise. Statistical analysis on the distortion of the interference pattern can then yield crucial information on the spatial correlation of the classical phases:
$C_{ij}=\langle e^{i\varphi_i}\,e^{-i\varphi_j}\rangle$, where $\langle\ldots \rangle$ stands for average over the whole lattice for a fixed distance between the sites.  Such a correlator is insensitive to the random variation of the phases from shot to shot, and can thus be enhanced by a further average over many runs of the experiment.  What do we expect on the quantitative behavior of this correlator?  

On the MI side of the phase transition, the correlator decays to zero as $C_{ij} \sim \exp(-|{\bf x}_i -{\bf x}_j|/\xi)$ for $|{\bf x}_i -{\bf x}_j| \geq \xi$.  For $|{\bf x}_i -{\bf x}_j| \leq \xi$, the correlator exhibits a power law decay typical for the critical point $\sim 1/|{\bf x}_i -{\bf x}_j|^{1+ \eta}$ with some universal exponent $\eta \approx 0.038$ 
in 2d lattice across the MI-SF transition and in a 3d gas undergoing the (finite temperature) BEC transition.   

On the SF side of the phase transition, the correlator retains a finite value $C_{\infty}$ characterizing the off diagonal long range order as $|{\bf x}_i -{\bf x}_j| \to \infty$.  This order parameter vanishes as a power law 
$C_{\infty}\sim |u-u_c|^{2\beta}$, with another universal exponent $\beta$ \cite{FSS}, as the critical point is approached.  The decay from $C_{ii}=1$ to $C_{\infty}<<1$ is described by the correlation length $\xi$ on the SF side. 

The critical point is practically reached when the correlation length exceeds the system size $L$.  In this case, the order parameter behaves as 
$C_{\infty}\sim L^{-2\beta/\nu}$ in accordance with the finite size scaling \cite{FSS}.  This corresponds to a power-law decay of the fringes with the system size $L$, which highly contrasts with the standard
interference situation of two coherent BEC clouds deep in the SF phase \cite{Ketterle}, where no $L$-dependence is to be anticipated.

\subsection{Density-density correlations in the interference pattern}

At long expansion times $t > \xi^2$, the phase domains start
overlapping so that visibility of the fringes with the reference BEC is
strongly reduced. However, effects of
the coherence of the initial many-body state can still be observed
in the density correlations, which can be measured 
by taking the average over many realizations \cite{Demler2}. Such experiments are matter-wave analog of the Hanbury-Brown and Twiss (HBT) effect
\cite{HBT} widely employed in quantum optics, and important
results in this direction have recently appeared~\cite{Bloch_noise,Aspect}. 

The density-density correlation function is defined by 
\begin{equation}
K({\bf x}_1,{\bf x}_2;t)=\langle n({\bf x}_1,t)n({\bf x}_2,t)\rangle,
\label{dens-dens}
\end{equation}
where the density operator is given by $n({\bf x},t)= \psi^\dagger ({\bf x},t) \psi({\bf x},t)$ with
\begin{equation}
\psi({\bf x},t)=\psi_l({\bf x},t)+\psi_{rf}({\bf x},t).
\label{field_exp}
\end{equation}
The subscripts $l$ and $rf$ refer to the states of the lattice-released and the reference clouds, respectively.  We assume as before  
\begin{equation}
\psi_{rf}({\bf x},t)=\frac{e^{i{\bf k}_r{\bf x}}}{\sqrt{\Omega}}\,b_{rf}
\label{ref_1}
\end{equation}
(a more practical setup will be discussed in section III).  
For the cloud released from the lattice, the operator $\psi_l({\bf x},t)$ is given by eq.(\ref{psi_l}).

The density operator can be split into three terms due to: 
lattice-released cloud $n_l=\psi_l^\dagger \psi_l$, the reference
cloud $n_{rf}=\psi_{rf}^\dagger\psi_{rf}$, and the interference between
the two $n_{int}=\psi_{rf}^\dagger\psi_l+\psi_l^\dagger\psi_{rf}$,
respectively. Then, among the many terms which result, we are mostly interested in the correlation function 
for the interference term $n_{int}$ alone. In the center-of-mass
${\bf r}=({\bf x}_1+{\bf x}_2)/2$ and relative ${\bf \rho}={\bf x}_1-{\bf x}_2$ coordinates, it reads: 
\begin{eqnarray}
K_{int}({\bf r},{\bf \rho};t)=
\langle n_{int}({\bf r}+{\bf \rho}/2)\,n_{int}({\bf r}-{\bf \rho}/2)\rangle=
\nonumber \\
n_{rf}\, e^{i{\bf k}_r\cdot\rho}\,\langle
\psi^\dagger_l({\bf r}+{\bf \rho}/2,t)\,\psi_l({\bf r}-{\bf \rho}/2,t) \rangle +c.c..
\label{K_int}
\end{eqnarray}
This term is directly proportional to the one-particle density
matrix in OL modified by free expansion. 
Below we will show that
the integration over the center of mass coordinate
${\bf r}$ results in the quantity
${\bar  K}_{int}({\bf \rho},t)=\int\!d{\bf r}\,K_{int}({\bf r},{\bf
\rho},t)$
which represents spatial correlations averaged
over the lattice and which turns out to be invariant in time and therefore insensitive to the expansion
\cite{noteniu2}.
In fact, ${\bar  K}_{int}$ represents the mean of the single particle shift operator
$\langle \exp(i{\bf \rho} {\bf p}_1)\rangle$, where ${\bf p}_1 $ stands for the momentum operator for one particle. During free expansion, the momentum of each particle is conserved, implying that the mean shift is a constant in time.

We comment on the strength of the remaining terms. The cross correlation between $n_{int}$ and 
$n_l+n_r$ vanishes upon ensemble average.
The remaining terms are from the correlator of $ n_l+n_{rf}$. The corresponding
terms produce either a uniform background or have a different spatial periodicity than $k_r$, and can thus be filtered out from an actual measurement by means of spatial Fourier analysis. 
The correlation function for the interference density can also be identified based on a strength analysis, if we have a relatively strong BEC for the reference cloud and a relatively low density for the lattice-released cloud.
(The latter scales as $1/t^3$ after expansion, which can be made weaker by waiting longer.)  In this case, it is even more convenient to consider the noise correlator \cite{Demler2}
$K'({\bf r},{\bf \rho},t)=\langle \Delta n({\bf r} + {\bf \rho}/2,t)\Delta n({\bf r} - {\bf \rho}/2,t)\rangle$,
which has no cross terms with respect to the three parts of the density operator.
The fluctuations $\Delta n\equiv n- \langle n \rangle$ will be suppressed in the reference cloud as long as the BEC is well formed.  The noise correlator for the lattice-released cloud is small (scales as $t^{-6}$) due to the assumed low density. Thus, the term (\ref{K_int}), which scales as $\sim n_{rf}t^{-3}$ turns out to be dominant.    
In our subsequent discussions we shall not explicitly distinguish $K'({\bf r},{\bf \rho},t)$ and $K_{int}({\bf r},{\bf \rho},t)$.

We can thus focus on the space-averaged correlation function at time $t=0$, which can be evaluated as 
\begin{equation}
{\bar  K}_{int}(\rho;t=0)=n_{rf}\,e^{i{\bf k}_r\rho}\,\sum_{ij}
\big\langle b_i^\dagger b_j \big\rangle\,e^{-\frac{1}{4a_w^2}(\rho-\rho_{ij})^2} +c.c..
\label{K_int_final}
\end{equation}
Here, we have assumed for simplicity a Gaussian form for the initial Wannier function:
\begin{equation}
W_0({\bf x})=a_w^{-3/2}\,e^{-x^2/2\,a_w^2}.
\label{W_0}
\end{equation}
The quantity $\big\langle b_i^\dagger b_j \big\rangle$ may be regarded as equal to the quasi-condensate density $\tilde n$ times the phase correlator $C_{ij}$ discussed in section IB.  Let $C(\rho)$ be an envelop function that equals the phase correlator at $\rho=\rho_{ij}$ and smoothly interpolates between the lattice vectors.  We expect this envelop to vary on the scale of the correlation length $\xi$, which is taken to be much bigger than the microscopic size of the Wannier function $a_w$.  Then we can write 
\begin{equation}
{\bar  K}_{int}(\rho;t=0)=n_{rf}\,e^{i{\bf k}_r\rho}\, C(\rho) \sum_{ij} \tilde n e^{-\frac{1}{4a_w^2}(\rho-\rho_{ij})^2}+c.c..
\label{K_int_final2}
\end{equation}
The phase correlator envelop of the lattice thus enters directly as a factor in the density-density correlation function.  The other important factor is the plane-wave modulation from the reference BEC. The factor from the lattice sum varies over the scale of the lattice spacing and can be easily separated out.  In the following analysis, we assume that a smoothening procedure has been applied to this factor, so that it is replaced by a number proportional to the total number $N$ of the particles in the lattice.  

Having considered the behavior of the phase correlator in Sec.\ref{IB}, we can now summarize our results for various regimes for the quantum MI-SF transition in d=2 OL and in d=3 traps undergoing finite temperature
BEC transition. As mentioned above, both systems are equivalent in terms of the universality of the transitions.
Here we assume that the condition of weak gradients \cite{troy} is satisfied in both cases. 
This condition can be formulated in terms of a single length $\sim L$ determining 
both the gradients and the system size, so that as $L$ is taken larger and larger the system becomes
progressively closer to the transition and the typical scaling  (with $L$) behavior can be observed.   
In both cases for $\xi<L$, in the phase were no symmetry is broken 
 \begin{eqnarray}
{\bar K}_{int}({\bf \rho},t)\sim n_{rf}N\frac{{\rm e}^{i{\bf k}_r{\bf \rho} - \rho/\xi}}{\rho^{1+\eta}} +c.c..
\label{K_xi}
\end{eqnarray}%
In what follows, we will call this phase {\it normal} without specifying if it refers
to the MI or to the thermal gas above the BEC transition.
When $\xi$ becomes larger than $L$, a system is in the critical regime, where ${\bar K}_{int}({\bf \rho},t)\sim e^{i{\bf k}_r\rho} /\rho^{1 + \eta}$.
The exponent $\eta$ \cite{FSS} is related to the exponents $\beta$ and $\nu$ by means of $2\beta /\nu = 1 +\eta$.  This implies, in particular, that measuring the magnitude of the correlator at $\rho \approx L$, which goes like $ L^{-2\beta /\nu}$, in progression of $L$ can provide valuable information on the exponent $\beta$.  These expressions are independent on particular details of a system undergoing the phase transition and are not distorted by the free expansion. 

In actual experiments, it is the columnar density that is usually measured in absorptive imaging.  In the case of a $3d$ BEC transition in a trap, the functional dependence in (\ref{K_xi}) is not changed much. If the columnar density is collected along $z-$axis which is normal to ${\bf k}_r$,
the 3d function (\ref{K_xi}) must be replaced by the integral
$ n_{rf}N{\rm e}^{i{\bf k}_r{\bf r}} \int dz {\rm e}^{- \sqrt{r^2+z^2}/\xi}/(r^2+z^2)^{(1+\eta)/2}$,
with ${\bf r} $ being 2d coordinates in the $xy$ plane.  As a function
of ${\bf r} /\xi$, this integral exhibits exponential behavior similar to eq.(\ref{K_xi}).
In the critical situation, with the correlator measured on distances $\sim L$,
the columnar density interference pattern will acquire one extra power of $L$
as $\sim L^{1 -2\beta /\nu}=L^{- \eta}$. 

The 2d MI-SF transition can naturally be realized in a stack of
identical and independent 2d lattices, and the columnar density can be
collected along the $z$-axis perpendicular to the layers, with
${\bf k}_r$ being along the layers.
Differently from the 3d case, this simply leads to a multiplicative
prefactor in (15), so that the correlator scales as $\sim L^{-2\beta /\nu}=L^{-(1+ \eta)}$.

Useful results can also be obtained from the Fourier transform 
${\cal Q}(\xi,L)= \int d{\bf \rho} {\bar K}_{int}({\bf \rho},t)\cos({\bf k}_r{\bf \rho})$.
In 2d OL, it must scale as 
\begin{eqnarray}
{\rm Normal} &:& \quad {\cal Q}(\xi,L) \sim N \xi^{d-1} =L^d\xi^{d-1} , 
\label{xi<L}  \\
{\rm Critical}&:&\quad {\cal Q}(\xi,L) \sim N L^{d-1 - \eta}=L^{2d -1- \eta}, 
\label{xi>L}\\
{\rm BEC}&:&\quad {\cal Q}(L) \sim N^2= L^{2d}, 
\label{Q_BEC}
\end{eqnarray}%
where the last line is given for the BEC case in which the phase correlator in (\ref{K_xi}) must be replaced by just a constant. The dimension $d$ should be understood as follows: $d=2$ in the MI-SF case (which can be realized as a stack of independent 2d-layers) and $d=3$ in the case of the BEC transition in a 3d trap.
As we will discuss later, these properties are quite robust with respect to
the clouds non-uniformities (provided these are weak in a sense discussed above). Thus, observing how the global characteristic 
${\cal Q}(\xi,L)$ scales with $L$ can provide direct information on criticality. 
More detailed discussion of the cases will be given later, where the practical issues such as finiteness and non-uniformity of the reference
cloud are considered.

\section{Non-destructive measurements of the space-time correlator}

A very important information about the system can be obtained if  the {\it in-situ} space-time
correlator ${\cal K}({\bf x}_2,\tau;{\bf x}_1,0)=\langle \psi_l^\dagger({\bf x}_2, \tau) \psi_l({\bf x}_1,0)\rangle$ is known, where $\tau$
refers to the time evolution inside the lattice or the trap. Specifically, the excitation spectrum
is provided by the $\tau-$dependence. Obviously, the destructive
scheme cannot yield such correlator -- the method described above
gives the one-particle density matrix, that is, the one-time correlator corresponding to
$\tau=0$.

A very flexible method of non-destructive detection \cite{non-des}
of the phase correlations has recently been
introduced. It relies on interference of the outcoupled atomic beams.
Another method of the space selective non-destructive imaging
has been employed in combination with the single atom detection \cite{ETHZ}.

Here, we show that the method of the noise spectroscopy applied
to the images of interference between two outcoupled atomic beams and two
reference BEC clouds with fixed relative phases
can provide the complete correlator  ${\cal K}({\bf x}_2,\tau; {\bf x}_1,0)$.
This statement can be illustrated in the frame of two successive measurements
of the interference contributions $A_1({\bf x}_1)= \psi^\dagger_{1rf}({\bf x}_1) \psi_o({\bf x}_1,0) + H.c.$
and $A_2({\bf x}_2)=\psi^\dagger_{2rf}({\bf x}_2) \psi_o({\bf x}_2,\tau) + H.c.$ taken at the time moments
$t=0$ and $t=\tau$, where $\psi^\dagger_{1rf}({\bf x}_1)$ and $\psi^\dagger_{2rf}({\bf x}_2)$ refer to the operators of the reference clouds
with well defined relative phases.
If the measurement is performed
rapidly enough, the field operators $\psi_o({\bf x}_1,0)$ and $\psi_o({\bf x}_2,\tau)$
describing the outcoupled atomic beams are replicas (modulo
multiplicative factors depending on the details of the extraction
scheme) of the field operators describing the lattice cloud $\psi_l({\bf x}_1,0)$ and $\psi_l({\bf x}_2,\tau)$, respectively.

Following the same logic applied to the case of the destructive imaging, as long as the
correlation length is large (includes many particles in the volume $\sim \xi^d$) one can
interpret the one-shot two interference images in terms of the phases of the two reference clouds $\varphi_{1rf},\,
\varphi_{2rf}$
and of the outcoupled beams $\varphi ({\bf x}_1,0),\, \varphi ({\bf x}_2,\tau)$.
Thus, $A_k({\bf x}_k)=\sqrt{n_{krf} n({\bf x}_k,t_k)}{\rm e}^{i(\varphi({\bf x}_k,t_k)-
\varphi_{krf}({\bf x}_k))} + c.c.$, where  $n_{krf}$ and $n({\bf x}_k,t_k)$ stand for the densities of the $k=1,2$ reference
clouds and of the outcoupled beams taken at the time moments $t_k$, respectively.

The single-shot interpretation can be extended into the correlation multi-shot analysis.
As the phases of the reference clouds are well defined with respect to
each other (this can be achieved by splitting one BEC), they can be taken out from the average describing the
multi-shot correlations:
 \begin{widetext}
\begin{eqnarray}\label{K_amp}
\langle A^*_2({\bf x}_2)A_1({\bf x}_1)\rangle=\sqrt{n_{1rf}n_{2rf}n({\bf x}_1,0)n({\bf x}_2,\tau)}
\langle {\rm e}^{i[\varphi_{2rf}({\bf x}_2)-\varphi_{1rf}({\bf x}_1) + \varphi({\bf x}_1,0)-\varphi({\bf x}_2,\tau)]}\rangle= {\cal A}({\bf x}_2, {\bf x}_1)
{\cal K}({\bf x}_2,\tau;{\bf x}_1,0),
\end{eqnarray}
\end{widetext}
where the multiplicative factor ${\cal A}({\bf x}_2, {\bf x}_1)=
\sqrt{n_{1rf}n_{2rf}}{\rm e}^{i(\varphi_{2rf}({\bf x}_2) -\varphi_{1rf}({\bf x}_1))}$ is due to the
reference clouds.
Expansion can be described in the same way as in the destructive case,
in particular it is easy to see that it does not affect the center of
mass average ${\bar {\cal K}}({\bf \rho},\tau)=\int d{\bf r} {\cal K}({\bf r} + {\rho }/2 ,\tau ;{\bf r} - {\rho }/2,0)$.
Thus, studying it as a function of the spatial and time separations will allow extracting
crucial information on the critical space-time behavior \cite{Sachdev}.


The above scheme can be extended to the case of higher order
correlator \cite{note2}, i.e. a $n$-body Green's function of the form
$K_n({\bf x}_1,t_1;\ldots ; {\bf x}_{2n},t_{2n})= \langle \Psi^\dagger({\bf x}_1,t_1)\ldots
\Psi^\dagger({\bf x}_n,t_n) \Psi({\bf x}_{n+1},t_{n+1}) \ldots \Psi({\bf x}_{2n},t_{2n})\rangle $
calculated at the points $({\bf x}_1,t_1); ... ({\bf x}_{2n},t_{2n})$ in space-time.
As $2n$ time points are involved, one needs to prepare $2n$
reference clouds with well defined relative classical phases $\varphi_{k,rf},\, k=1,2...2n$,
and to outcouple
$2n$ atomic beams at the corresponding time moments $t_1, t_2 ... t_{2n}$.
Each atomic beam should be allowed to interfere with a corresponding reference cloud,
so that single-shot images give the amplitudes $A_k({\bf x}_k),\, k=1,2,\ldots, 2n$. Then, the multi-shot
correlation analysis provides the correlator $K_n({\bf x}_1,t_1;\ldots ;
{\bf x}_{2n},t_{2n})$.

\section{Practical issues}\label{prac}

In the previous discussion we have adopted a simplified approach which ignores
non-uniformity of a reference cloud as well its expansion. In fact, under these conditions,
it is hard to achieve statistical independence of the both clouds. In reality, a reference
cloud of some finite size $R_o > L$ should be created some distance $x_o > R_o$ away from
the OL so that there is, practically, no overlap or it is exponentially small so that
it can be considered as zero. 

\begin{figure}[h]
\begin{center}
\epsfxsize=8.5cm
\epsfbox{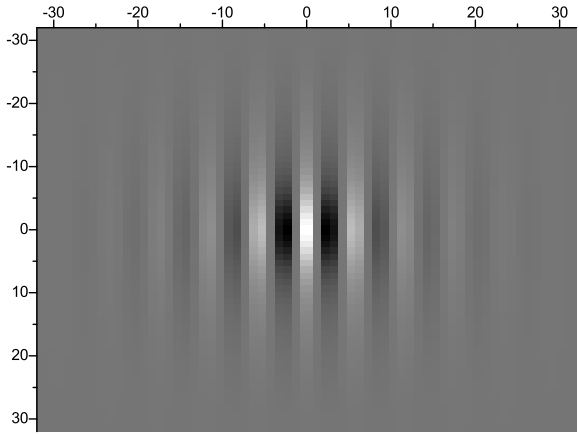}
\caption{
\label{contour}
A typical interference density distribution for the noise correlator $K'({\bf \rho},t)$
 (scaled in arbitrary units)
for a stack of 2d OL layers of size 32$\times$32 (unit cells) as given by eqs.(\ref{K_int_gen}, \ref{Sij}, \ref{phi_ij}). 
The correlator $\langle b^\dagger_ib_j\rangle$ is taken in the form (\ref{bi},\ref{bi_G}),
with $\xi \to \infty$. }
\end{center}
\end{figure}

This implies that $\int d^d\rho {\bar K}_{int}(\rho;t)=0$ for any $t$ because
it is simply proportional to the quantum mechanical overlap of the one-particle
wavefunctions of the OL and the reference cloud, which, initially, was taken
as zero (or exponentially small). [The form
(\ref{K_xi}) does not satisfy this condition due to the approximation
of infinite size of the reference cloud.] This can be seen explicitly
by noting that $\int d^d\rho {\bar K}_{int}(\rho;t)=
\int d{\bf x}_1\int d{\bf x}_2=\langle n({\bf x}_1,t)n({\bf x}_2,t)\rangle$
and that the integral of the total density $\int d{\bf x} n({\bf x},t)=
\int d{\bf x} \psi^\dagger_l({\bf x},t)\psi_l({\bf x},t) + \int d{\bf x} \psi^\dagger_r({\bf x},t)\psi_r({\bf x},t)$
because the interference part $\int d{\bf x} \psi^\dagger_l({\bf x},t)\psi_r({\bf x},t)+ H.c.$ 
is essentially zero due to orthogonality (or negligible overlap) of the one-particle
states of the tested and reference clouds.
However, the Fourier harmonic 
${\cal Q}(\xi,L,t)= \int d^d\rho {\bar K}_{int}(\rho;t)\cos({\bf k}_r {\bf \rho})$
is not an integral
of motion and does depend on time, reaching its maximum at the closest approach of the centers
of mass of the two clouds. 

Here we shall consider a situation
when both clouds are released the distance $x_o$ apart at $t=0$ with some relative
momentum ${\bf k}_r$ toward each other to achieve maximum overlap of the centers
of mass at the time
moment $t=x_o/k_r$ (in atomic units $m=1, \hbar=1$). During the free flight period,
the expansion due to quantum pressure deforms the clouds and the resulting image.
Accordingly, the correlator ${\bar K}_{int}(\rho;t)$ cannot be considered
as a constant of motion anymore, and the time-dependence must be calculated explicitly. 
Further image analysis should be based on the fit procedure discussed below.

\begin{figure}[h]
\begin{center}
\epsfxsize=8.0cm
\epsfbox{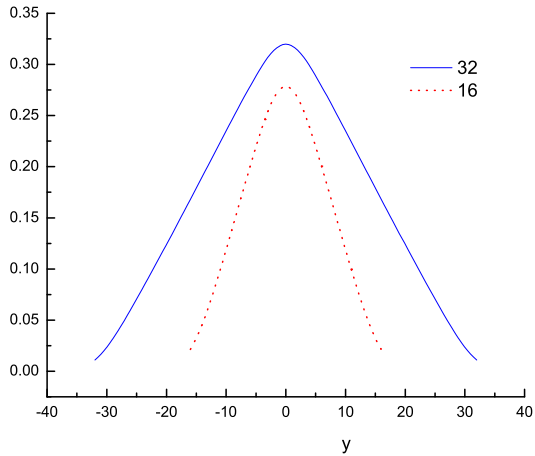}
\caption{
\label{BEC_y}(color online)
The profile $K'(\rho_x=0, \rho_y=y,t)$ (arbitrary units) for a stack of 2d OL layers in the BEC state for two sizes $L=16,32$.
The horizontal axis is the OL y-coordinate in units of $a_l$.
The typical pyramidal shape is due to the non-uniformity of the OL cloud.}
\end{center}
\end{figure}

\begin{figure}[h]
\begin{center}
\epsfxsize=8.0cm
\epsfbox{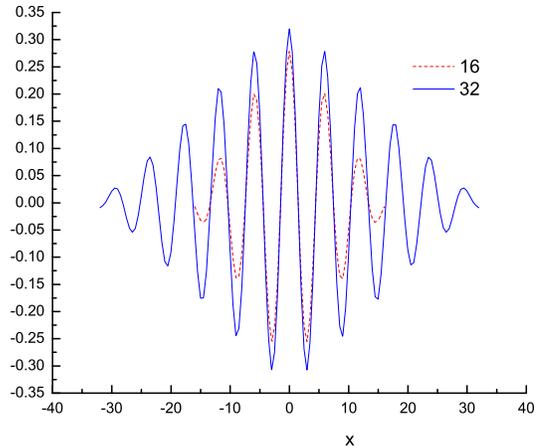}
\caption{
\label{BEC_x}(color online)
The same system as in FIG.~\ref{BEC_y}, with the profile taken along $\rho_x=x, \rho_y=0$. 
}
\end{center}
\end{figure}

Instead of the simplified form (\ref{ref_1}), we now take
the reference cloud wavefunction at $t=0$ as
\begin{equation}
\psi_{rf}({\bf x},t=0)=\frac{1}{(\sqrt{\pi}R_o)^{3/2}}
e^{i{\bf k}_r({\bf x} -{\bf x}_o) - \frac{({\bf x} - {\bf x}_o)^2}{2R_o^2}}\,b_{rf},
\label{ref_2}
\end{equation}
where ${\bf x}_o$ stands for its initial position.
Then, after time $t>0$ of free expansion it evolves into
\begin{equation}
\psi_{rf}({\bf x},t)=\frac{R_o^{3/2}}{\pi^{3/4} (R^2_o + it)^{3/2}}
e^{i{\bf k}_r({\bf x} -{\bf x}_o(t)) - \frac{({\bf x} - {\bf x}_o(t))^2}{2(R_o^2 +it)}}\,b_{rf},
\label{ref_22}
\end{equation}
where ${\bf x}_o(t)={\bf x}_o+{\bf k}_r t$.
This form must be used in eq.(\ref{field_exp}) together with eq.(\ref{psi_l}) and
eq.(\ref{t>t_1}) in the limit $R_o \gg t/R_o \gg a_w$. Then, integrating over the center
of mass coordinate, as was done in derivation of eq.(\ref{K_xi}), we obtain the

correlator
\begin{equation}
{\bar  K}_{int}({\bf \rho};t)=\frac{n_{rf}}{t^3} \,\sum_{ij}
\big\langle b_i^\dagger b_j \big\rangle\,{\rm e}^{S_{ij}}\cos(\varphi_{ij})\, ,
\label{K_int_gen}
\end{equation}
where the following notations are introduced
\begin{eqnarray}
S_{ij}&=& - \frac{R^2_o}{4t^2}({\bf \rho}-{\bf \rho}_{ij})^2 ,
\label{Sij} \\
\varphi_{ij}&=& {\bf k}_r {\bf \rho} + \frac{1}{t}({\bf R}_{ij} +{\bf x}_o - t{\bf k}_r)({\bf \rho}-{\bf \rho}_{ij}),
\label{phi_ij}
\end{eqnarray}
and ${\bf R}_{ij}=({\bf x}_i + {\bf x}_j)/2$ (the OL
spacing we take as a unit of distance), with the initial positions ${\bf r}_i$ of the lattice
sites counted from the origin located in the middle of the lattice. The overall numerical
factor in eq.(\ref{K_int_gen}) is omitted. 
For 
demonstrative purposes we consider a model form of the lattice correlator on the insulator
side of the transition as
\begin{eqnarray}
\langle b_i^\dagger b_j \rangle&\sim&\frac{\exp(-G_{ij})}{|{\bf x}_i -{\bf x}_j|^{1+\eta}}   
\label{bi} \\
G_{ij} &=&\frac{x_i^2 + x_j^2}{L^2} + \frac{|{\bf x}_i -{\bf x}_j|}{\xi},
\label{bi_G}
\end{eqnarray}
where the term $\frac{x_i^2 + x_j^2}{L^2}$ reflects the non-uniformity in the trap and
$\eta \approx 0.04$.
In the BEC state:
\begin{eqnarray}
\langle b_i^\dagger b_j \rangle&\sim&\exp(-\frac{x_i^2 + x_j^2}{L^2}).   
\label{bi_BEC}
\end{eqnarray}
The spatial profile of the correlator ${\bar  K}_{int}({\bf \rho};t)$ is presented on 2d columnar density
plot FIG.~\ref{contour} for the
case $\xi \gg L$ in eqs.(\ref{bi},\ref{bi_G}) and for the moment of the maximum overlap
$t=x_o/k_r$. Typical profiles of the noise correlator for the BEC case are shown on
FIG.~\ref{BEC_y} and FIG.~\ref{BEC_x}.


The integral ${\cal Q}(\xi,L)$ becomes
\begin{eqnarray}
{\cal  Q}(\xi,L)=\int d^d \rho{\bar  K}_{int}({\bf \rho};t)\cos({\bf k}_r{\bf \rho}) \sim 
\nonumber \\
\sum_{ij}
\big\langle b_i^\dagger b_j \big\rangle\,{\rm e}^{-({\bf R}_{ij} +{\bf x}_o - t{\bf k}_r)^2/R^2_o}
\label{Q_xiL}
\end{eqnarray}

The contribution $ \sim ({\bf R}_{ij} +{\bf x}_o - t{\bf k}_r)({\bf \rho}-{\bf \rho}_{ij})$ to
the phase factor $\varphi_{ij}$ in eq.(\ref{K_int_gen}) can destroy the correlations
imprinted in $\big\langle b_i^\dagger b_j \big\rangle$. However, if $R_o \gg L$, then,
at the maximum overlap of the centers of mass
of both clouds (at $t=x_o/k_r$) this term can be estimated
as $\approx L/R_o \ll 1$, and,thus, becomes insignificant. In what follows we choose the following
parameters $R_o= 4L,\, x_o=8.5L,\, t=8L$. Accordingly, $k_r=x_o/t \approx 1.1$ , so that the wavelength $2\pi/k_r \approx 6$ in units of the lattice spacing.

\begin{figure}[h]
\begin{center}
\epsfxsize=8.0cm
\epsfbox{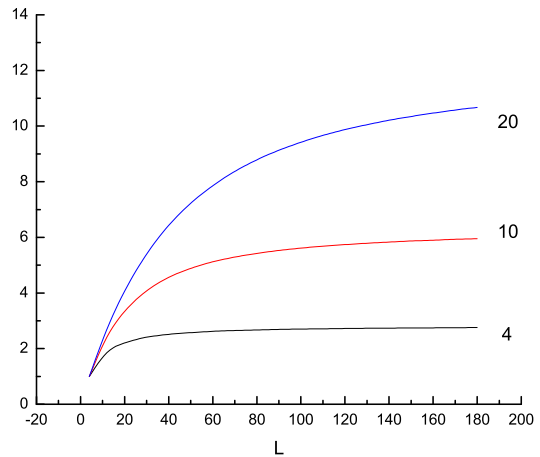}
\caption{
\label{Q_xi<L}(color online)
The integral ${\cal Q}(\xi,L)/L^d,\, d=2$ as a function of the system size $L$ (in units of $a_l$) for three values of the 
correlation length $\xi=4,10,20$ shown on the graph. The curves are normalized by ${\cal Q}(\xi,L=4)/4^2$.}
\end{center}
\end{figure}

\begin{figure}[h]
\begin{center}
\epsfxsize=8.0cm
\epsfbox{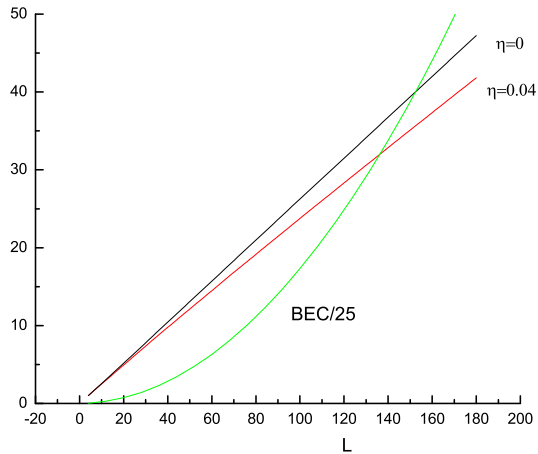}
\caption{
\label{Q_crit} (color online)
The integral ${\cal Q}(\xi,L)$ normalized by
${\cal Q}(\xi,L=4)$ as a function of the system size $L$ (in units of $a_l$) for three cases: 1) labeled as
$\eta=0$, represents the correlator (\ref{bi},\ref{bi_G}) with $\eta=0$; 2) labeled as
$\eta=0.04$, the same as in the case 1) except that $\eta=0.04$; 3) labeled as BEC/25, 
corresponds to the BEC state as given by eq.(\ref{bi_BEC}) and reduced by the factor of 25 to
fit into the plot.}
\end{center}
\end{figure}

The quantity ${\cal Q}(\xi,L)$ is plotted for the case (\ref{xi<L}) in FIG.~\ref{Q_xi<L}.
On FIG.~\ref{Q_crit}, the quantity ${\cal Q}(\xi,L)/L^d,\, d=2$ is represented
for the critical correlators ($\xi =\infty$) )(\ref{bi}) for two values
of the critical exponent $\eta=0$ and $\eta \approx 0.04$ in order to demonstrate
how a small value of $\eta$ can be resolved. Direct fits of these dependencies from eq.(\ref{xi>L})
give $\eta=0.00 +- 0.001$ for the case $\eta=0$ and $\eta=0.036 +- 0.001$ for the case $\eta=0.04$.
For a comparison, the case corresponding to the BEC is presented as well. Thus, this graph clearly
demonstrates the drastic differences between critical and fully coherent clouds. Obviously,
the scaling dependences are not obscured by the clouds non-uniformity because a single
size $L \ll \xi$ determines the lattice size, the dimension of the reference cloud $R_o \sim L$ and
the extent of the non-uniformity as implied by eq.(\ref{bi_G}).

 \section{Conclusion and acknowledgment}
We have proposed a method for detecting
critical space-time correlations in the vicinity of phase transitions in optical lattices and traps
within the concept of the finite size scaling.
Its essential component is the noise-correlation analysis of the pattern obtained from the interference
with some reference BEC cloud. The interference term turns out to be, on one hand, sensitive to the diverging critical correlations, and, on the other hand,
essentially insensitive to the underlying structure
of a lattice, so that this method can be also used to study the critical behavior
in lattices with imposed disorder. The finite size scaling concept can be
effectively applied for extracting the critical exponents from the integral
quantity ${\cal Q}(\xi,L)$.
The non-destructive scheme employing several reference clouds with well defined relative phases
is suggested as a tool for obtaining many-body, multi-time correlators.

One of us (A.B.K.) acknowledges useful discussions with B.V. Svistunov and with the members of the ETHZ Quantum Optics Group: T. Donner, T. Esslinger, K. Guenter, M.K\"ohl, H. Moritz, A. \"Ottl and T. Stoeferle.
This work was  supported by NSF (PHY-0426814 and PHY-0244688), PSC-CUNY (66556-0036), and the Welch Foundation.
The authors also acknowledge hospitality of the Aspen Center for Physics during the Summer 2005 Workshop on ultra-cold atoms.



\begin{thebibliography}{99}
\bibitem{MI-SF} M. Greiner,  O. Mandel, T. W. Hansch, and I. Bloch,
Nature {\bf 415}, 39(2002)

\bibitem{Zhou} E. Demler and F. Zhou
Phys. Rev. Lett. {\bf 88}, 163001(2002)

\bibitem{SCF} A.B. Kuklov and B.V. Svistunov
Phys. Rev. Lett. {\bf 90}, 100401(2003)


\bibitem{We} A.B. Kuklov, N.V. Prokof'ev and B.V. Svistunov
Phys. Rev. Lett. {\bf 92}, 030403(2004); A.B.Kuklov, N.V. Prokof'ev and B.V. Svistunov,
Phys. Rev. Lett. {\bf 92}, 050402(2004).


\bibitem{Demler} L.-M. Duan, E. Demler, and M.D. Lukin
Phys. Rev. Lett. {\bf 91}, 090402(2003).

\bibitem{Boseglass}
N.Prokof'ev and B. Svistunov, Phys.Rev.Lett. {\bf 92}, 015703 (2004);
K.G.Balabanyan,  N.Prokof'ev and B. Svistunov, Phys.Rev.Lett. {\bf 95}, 055701 (2005).

\bibitem{glass_exp}
B.Damski, J.Zakrzewski, L.Santos, P.Zoller and M.Lewenstein, Phys.Rev.Lett. {\bf 91}, 080403 (2003).

\bibitem{Ketterle}
M.R. Andrews, C.G. Townsend, H.-J. Miesner, D.S. Durfee,
D.M. Kurn, and W. Ketterle, Science {\bf 275}, 637-641 (1997).

\bibitem{note}
J. Stenger, S. Inouye, A. P. Chikkatur, D. M. Stamper-Kurn, D. E. Pritchard, and W. Ketterle,
Phys.Rev.Lett. {\bf 82},4569 (1999)


\bibitem{Bloch1}
I. Bloch, T. W. Haensch and T. Esslinger, Nature {\bf 403}, 166 (2000). 

\bibitem{Damle}
K. Damle, T. Senthil, S. N. Majumdar, S. Sachdev , Europhysics Letters, {\bf 36}, 7 (1996).

\bibitem{troy2}
S. Wessel, F. Alet,M. Troyer, and G. G. Batrouni, Phys. Rev. {\bf A 70}, 053615 (2004)

\bibitem{troy}
O. Gygi, H. G. Katzgraber, M. Troyer, S. Wessel, G. G. Batrouni, cond-mat/0603319.





\bibitem{FSS}
D.J.Amit and V.Martin-Mayor, {\it Field Theory, the Renormalization Group, and Critical Phenomena},
World Scientific, 2005.

\bibitem{imaging}

F. Gerbier, A. Widera, S. Foelling, O. Mandel, T. Gericke, I. Bloch,
Phys.Rev.Lett. {\bf 95}, 050404 (2005).


\bibitem{Borya_Kolya}
V.A. Kashurnikov, N.V. Prokof'ev, and B.V. Svistunov, Phys. Rev. A {\bf 66}, 031601(R) (2002).
\bibitem{Popov}
V.N. Popov, {\it Functional integrals in quantum field theory and statistical physics},
Reidel, Dordrecht, 1983.
\bibitem{Kagan}
Yu. Kagan, B.V. Svistunov, G.V. Shlyapnikov, Sov. Phys. JETP {\bf 66}, 314 (1987).
\bibitem{Borya}
N. Prokof'ev, O. Ruebenacker, and B. Svistunov
Phys. Rev. {\bf A69}, 053625 (2004).
\bibitem{Sachdev}
S. Sachdev, {\it Quantum phase transitions}, Cambridge, 1999.
\bibitem{note1}
The criticality of the quantum MI-SF transition in $d=3$ can well be described by the
mean field approach in $D=4$ \cite{Sachdev}.
\bibitem{alpha_exp}
J. A. Lipa, J. A. Nissen, D. A. Stricker, D. R. Swanson, and T. C. P. Chui,
Phys.Rev. {\bf B 68}, 174518 (2003)
\bibitem{Burovski}
M. Campostrini, M. Hasenbusch, A. Pelissetto, P. Rossi, and E. Vicari,
Phys.Rev. {\bf B 63}, 214503 (2001);
E. Burovski, J. Machta, N.V. Prokof'ev, B.V. Svistunov,
cond-mat/0507352.


\bibitem{Altman}
E.Altman and A.Auerbach,  Phys. Rev. Lett. {\bf 89}, 250404(2002).




\bibitem{WallsMilburn} D. F. Walls and G. J. Milburn, {\em Quantum
  Optics} (Springer, Berlin, 1994).

\bibitem{vortex}
Eric L. Bolda and Dan F. Walls, Phys. Rev. Lett. {\bf 81}, 5477 (1998);
J. Tempere and J. T. Devreese, Solid State Commun. {\bf 108}, 993 (1998);
S. Inouye, S. Gupta, T. Rosenband, A. P. Chikkatur, A. Gorlitz, 
T. L. Gustavson, A. E. Leanhardt, D. E. Pritchard, and W. Ketterle, 
Phys. Rev. Lett. {\bf 87}, 080402 (2001);
F. Chevy, K. W. Madison, V. Bretin, and J. Dalibard, 
Phys. Rev. A {\bf 64}, 031601(R) (2001);




\bibitem{HBT} R.Hanbury Brown, R.Q.Twiss, Nature {\bf 178}, 1046 (1956)

\bibitem{Bloch_noise}
S.F\"olling, F.Gerbier, A.Widera, O. Mandel, T. Gericke \& I.Bloch,
Nature {\bf 434}, 481 (2005)
\bibitem{Demler2}
E. Altman, E. Demler, M. D. Lukin,
Phys. Rev. {\bf A 70}, 013603 (2004)
\bibitem{Aspect}
 M. Schellekens, R. Hoppeler, A. Perrin, J. Viana Gomes, D. Boiron, A. Aspect, C. I. Westbrook, cond-mat/0508466;
 J. Esteve, J.B. Trebbia, T. Schumm, A. Aspect, C. Westbrook, I. Bouchoule, cond-mat/0510397.
\bibitem{IC_YC} I. Carusotto and Y. Castin, Phys. Rev. Lett. {\bf 94},
223202 (2005).
\bibitem{Polkovnikov}
A. Polkovnikov, E. Altman, E. Demler,
cond-mat/0511675.


\bibitem{YC_JD} For a discussion of the use of coherent vs. Fock states
for a description of the BEC phase, see e.g.: Y. Castin and J. Dalibard,
Phys. Rev. A 55, 4330–4337 (1997).

\bibitem{noteniu} The amplitude is defined by $A({\bf x})=\langle S|b^\dagger_{rf} \psi_l |S\rangle /\sqrt{\Omega} $, which may be viewed as a {\it projection of the lattice field operator $\psi_l$ on the coherent states} of the reference cloud. This implies a possibility of imaging of the {\it wavefunction} (not its density!) of the lattice atoms as long as the reference cloud can be characterized by a robust well defined classical phase.

\bibitem{noteniu2} This statement can be proven by the explicit
calculation of the
evolution of the density-density correlator as
${\bar
  K}_{int}(\rho;t)=n_r\, e^{i{\bf k}_r\cdot\rho}\,
\int\!d{\bf r}\,d{\bf y}\,d{\bf y}'\,G_t^*({\bf r}+\rho/2-{\bf y})\\
G_t({\bf r}-\rho/2-{\bf y}')\,\big\langle \psi^\dagger_l({\bf y},0)\,\psi_l({\bf y}',0)
\big\rangle$,
with $G_t({\bf x})$ being the free-particle propagator.
Using the unitarity of $G_t({\bf x})$:
$\int\! d{\bf x}\,G_t^*({\bf x}+{\bf y})\,G_t({\bf x}+{\bf y}')=\delta({\bf y}-{\bf y}')$
as well as uniformity of space
and exchanging the order of integration, one
immediately finds that ${\bar K}_{int}({\bf \rho},t)$ is a constant of motion.
It is important to note that, while the independence of the integrated one-particle correlator
$\int d{\bf r} \langle \psi^\dagger_l({\bf r} + {\bf \rho}/2,t)\psi_l({\bf r} - {\bf \rho}/2,t)\rangle$
on the free expansion time $t$ is a sole consequence of the uniformity and isotropy of space,
the time independence of the two-body correlator (\ref{K_int}) integrated over the center of mass coordinate 
requires also a specific form (\ref{ref_1}) for the reference cloud wavefunction. As it is discussed
in the Sec. \ref{prac}, finite dimensions of the reference cloud affect the above result.


\bibitem{non-des}
M. Saba, T. A. Pasquini, C. Sanner, Y. Shin, W. Ketterle, D. E. Pritchard,
Science {\bf 307}, 1945 (2005);
Y. Shin, G.-B. Jo, M. Saba, T. A. Pasquini, W. Ketterle, and D. E. Pritchard,
Phys. Rev. Lett. {\bf 95}, 170402 (2005).

\bibitem{ETHZ}
T.Bourdel, T. Donner, S. Ritter, A. \"Ottl, M.K\"ohl,
and T. Esslinger, cond-mat/0511234

\bibitem{note2}
It is worth noting that performing the correlation analysis of the image from one outcoupled
beam, say, number 1, allows obtaining (in a non-destructive manner!) the same result
as (\ref{K_int_final2}),(\ref{K_xi}) for the one-time correlator discussed in the frame of the
destructive measurements. In other words,
$\langle A^*_1({\bf x})A_1({\bf y})\rangle=n_{1rf}{\rm e}^{i(\varphi_{1rf}({\bf x}) -\varphi_{1rf}({\bf y}))}\,\,
{\cal K}({\bf x},0;{\bf y},0).
$




\end{thebibliography}
\end{document}